\documentclass[referee]{raa}            

\usepackage{graphicx,times}             
\usepackage{threeparttable}
\usepackage{natbib}
\usepackage{amssymb,amsmath}
\bibpunct{(}{)}{;}{a}{}{,}

\usepackage[pagebackref=true]{hyperref}

\begin{document}

  \title{Asteroseismology study of a new faint ZZ Ceti J053009.62+594557.0 discovered in WFST
}

   \volnopage{Vol.0 (20xx) No.0, 000--000}      
   \setcounter{page}{1}          

   \author{Yonghui Yang 
      \inst{1}
   \and Jincheng Guo
      \inst{1}
    \and Jie Lin
      \inst{2,3}
      \and Tinggui Wang
      \inst{2,3,4}
      \and Ning Jiang
      \inst{2,3}
      \and Yibo Wang
      \inst{2,3}
      \and Lulu Fan
      \inst{2,3,4}
      \and Min Fang
      \inst{5}
      \and Bin Li
      \inst{5}
      \and Feng Li
      \inst{6}
      \and Hao Liu
      \inst{6}
      \and Ming Liang
      \inst{7}
      \and Wentao Luo
      \inst{4}
      \and Jinlong Tang
      \inst{8}
      \and Hairen Wang
      \inst{5}
      \and Jian Wang
      \inst{6,4}
      \and Yongquan Xue
      \inst{2,3}
      \and Dazhi Yao
      \inst{5}
   \and Hongfei Zhang
      \inst{6}
   }

   \institute{Department of Scientific Research, Beijing Planetarium, 100044, Beijing, People’s Republic of China; {\it andrewbooksatnaoc@gmail.com (JC.Guo)}\\
        \and
             Department of Astronomy, University of Science and Technology of China, Hefei 230026, People's Republic of China; {\it linjie2019@ustc.edu.cn (J.Lin)}\\
        \and
             School of Astronomy and Space Sciences, University of Science and Technology of China, Hefei, 230026, People's Republic of China\\
             \and
             Institute of Deep Space Sciences, Deep Space Exploration Laboratory, Hefei 230026, China\\
             \and
             Purple Mountain Observatory, Chinese Academy of Sciences, Nanjing 210023, China\\
             \and
             State Key Laboratory of Particle Detection and Electronics, University of Science and Technology of China, Hefei 230026, China\\
             \and
             National Optical Astronomy Observatory (NSF’s National Optical-Infrared Astronomy Research Laboratory) 950 N Cherry Ave. Tucson Arizona 85726, USA\\
             \and
             Institute of Optics and Electronics, Chinese Academy of Sciences, Chengdu 610209, China\\
\vs\no
   {\small Received 20xx month day; accepted 20xx month day}}

\abstract{ In this work, we present a detailed asteroseismological analysis of WFST J053009.62+594557.0, a newly discovered faint pulsating white dwarf by the Wide Field Survey Telescope (WFST) with a Gaia G magnitude of 19.13. Analysis of two nights of high-precision WFST $g$ band photometry reveals three significant pulsation frequencies with high signal-to-noise ratios. Follow-up P200/DBSP spectroscopy classifies the object as a DA white dwarf with $T_{\rm eff}$=11,609 $\pm$ 605\,K and $M=0.63 \pm$ 0.22\,${M_{\odot}}$. To probe its internal structure, we construct asteroseismological models with the \texttt{White Dwarf Evolution Code (WDEC)}. After exploring sufficient matching models, best-fitting solutions yield $T_{\rm eff}=11,850 \pm 10$ K and $M=0.600 \pm 0.005\,{M_{\odot}}$, consistent with independent constraints from Gaia color–magnitude diagram, Gaia XP spectrum, P200 spectral fitting, SED fitting, and Gaia parallax. It has shown that the asteroseismological distance agrees with the Gaia parallax to 1.45\%. }


   \authorrunning{Yang, Guo, Lin  et al. }           
   \titlerunning{A new faint ZZ Ceti discovered in WFST}  

   \maketitle

%
%
\section{Introduction}           
\label{sect:intro}

In the Milky Way galaxy, approximately 98\% of low- to intermediate-mass stars (with masses under 10 or 11 $M_{\odot}$ ) will evolve into white dwarfs (WDs) during their late evolutionary phases \citep{2008ARA&A..46..157W,2015ApJ...810...34W,2018MNRAS.480.1547L,2022PhR...988....1S}. These WDs have an average mass of roughly 0.6 $M_{\odot}$, radii of about 0.01 $R_{\odot}$, and densities on the order of $10^{6}$ $g/cm^{3}$ \citep{2016MNRAS.461.2100T,2020FrASS...7...47C}. These physical characteristics make them small and dense, thus ideal natural laboratories for studying extreme physical conditions. As WDs no longer undergo core nuclear fusion, their evolution is primarily characterized by cooling. Once a WD enters the pulsating instability strip, it can be classified into three pulsating WD categories based on its atmospheric composition: DAV (hydrogen-rich atmosphere), DBV (helium-rich atmosphere), and DOV (hot helium-rich atmosphere) \citep{2021A&A...651A..14B,2022FrASS...9.9045G,2022ARA&A..60...31K}. 

Recent high-cadence surveys like Kepler \citep{2010Sci...327..977B}, K2 \citep{2014PASP..126..398H} and TESS \citep[Transiting Exoplanet Survey Satellite,][]{2015JATIS...1a4003R} have provided abundant high-precision, data-rich pulsating WD samples, which strongly enrich asteroseismology research. However, among the over 500 DAVs identified to date \citep{2017ApJS..232...23H,2021A&A...651A..14B,2019A&ARv..27....7C,2021ApJ...912..125G,2022MNRAS.511.1574R,2025ApJ...984..112R,2020AJ....160..252V}, only a small proportion have undergone detailed asteroseismological studies. These stars are typically identified by their short periods ranging from 100\,s to 1500\,s and low amplitudes of around 1.0\% \citep{2023A&A...669A..62B}. 
The pulsations of DAV stars are characterized by nonradial g-modes, where gravity is the restoring force. This pulsation mode is driven by a mechanism closely associated with the partial ionization of hydrogen in the outer hydrogen-rich atmosphere of DAV stars \citep{1982PhDT........27W,2008ARA&A..46..157W,2008PASP..120.1043F,2010A&ARv..18..471A}. Specifically, the observed pulsations are driven by a combination of the $\kappa - \gamma$ mechanism \citep{1981A&A...102..375D,1982ApJ...252L..65W} and convective processes \citep{1991MNRAS.251..673B,1999ApJ...511..904G}. For a more comprehensive understanding of the theory and observations related to DAV stars, we refer to several review papers and their references therein \citep{2008ARA&A..46..157W,2008PASP..120.1043F,2019A&ARv..27....7C,2020FrASS...7...47C}.

Asteroseismology allows us to investigate the internal structure and chemical composition of pulsating WDs \citep{2008ARA&A..46..157W,2010A&ARv..18..471A,2019A&ARv..27....7C}. This is achieved by constructing theoretical models and comparing their predicted oscillation periods with observed ones. Currently, two main methods are used in the asteroseismological analysis of pulsating WDs: fully evolutionary models \citep{2010ApJ...717..897A,2010ApJ...717..183R,2025ApJ...984..112R,2017A&A...599A..21D,2024A&A...686A.140C} and parametrized models \citep{2009MNRAS.396.1709C,2013MNRAS.429.1585F,2014ApJ...794...39B,2018Natur.554...73G,2021ApJ...922..138G,2024MNRAS.528.5242G,2023A&A...669A..62B} . Fully evolutionary models track a star's evolution through stages like the main sequence and red-giant phases to its final WD stage. They help explore the detailed physical mechanisms of each evolutionary phase. In contrast, parametrized models do not involve specific evolutionary processes. They built WD models by varying input parameters, with the parameterized chemical profile offering a more extensive exploration space to identify the optimal asteroseismological model.

This paper presents an asteroseismological analysis of a newly discovered faint DAV star, WFST J053009.62 + 595557.0 (hereafter WFST J0530). The star's J2000 coordinates are R.A. = 05:30:09.63 and Dec. = +59:45:57.88. WFST J0530 was first appeared in Gaia DR2 data in 2018 \citep{2018MNRAS.480.4505J} and later reported in Gaia DR3 data \citep{2021MNRAS.508.3877G}. In 2024, based on the Gaia XP spectrum, \cite{2024A&A...682A...5V} determined the star's spectral parameters as $T_{\rm eff}$ = 11,657 $\pm$ 505\,K, $M$ = 0.619 $\pm$ 0.070\,$M_{\odot}$ and log\,$g$ = 8.025  $\pm$ 0.086. Previously, we reported WFST J0530 for the first time in \cite{Lin+etal+2025+WFST} and revealed the presence of a periodic signal of 6.7 minutes.

This paper is organized as follows. Photometric and spectroscopic observations are described in Section 2. The asteroseismological analysis is presented in Section 3. Detailed discussions are provided in Section 4. Finally, conclusions and summaries are given in Section 5.

\section{Observation}

\subsection{Gaia}

Gaia’s all-sky survey provides precise astrometric and photometric measurements for WFST J0530. Its G-band magnitude is 19.1347$\pm$0.0037 \citep{2023A&A...674A..33G}. The proper motions are $\mu_{\alpha}$ = -0.634$\pm$0.192\,mas\,yr$^{-1}$ and $\mu_{\delta}$ = 3.094$\pm$0.184\,mas\,yr$^{-1}$ \citep{2020yCat.1350....0G}. The source has a Gaia trigonometric parallax of 3.4973$\pm$0.2282 mas, corresponding to a geometric distance of 285.935$\pm$18.657 pc \citep{2020yCat.1350....0G}. With this distance, the apparent magnitude G is directly converted into an absolute magnitude. Meanwhile, Gaia’s three-band photometry (G, BP, and RP) delivers a high-precision (BP $-$ RP) color; once corrected for extinction, this places the WFST J0530 at an accurate location in the color–magnitude diagram \citep{2016A&A...595A...1G}. Based on this location and combined with WD evolutionary tracks, the effective temperature and mass of WFST J0530 can be inferred. Furthermore, Gaia’s XP spectra also yield estimates of effective temperature, surface gravity, and mass for WFST J0530, offering a reference for the outcomes of asteroseismological modeling \citep{2024A&A...682A...5V}.

\subsection{WFST}
WFST located on the Saishiteng Mountain of Lenghu at an altitude of 4200 meters \citep{Deng+etal+Nature+lenghu}. It was originally designed to conduct deep-sky surveys of the northern hemisphere. Compared with ZTF \citep[Zwicky Transient Facility,][]{2019PASP..131a8002B}, its survey depth can be extended by two magnitudes towards fainter stars \citep{Lei+etal+2023+limiting}. A 2.5-meter optical mirror is equipped on it, with a field of view of 6.5 square degrees \citep{Lou+etal+2016+WFST+design}. High-resolution imaging is able to provide high-quality data for astronomical research \citep{2023SCPMA..6609512W}.

During its six-year observation program, WFST will collect vast amounts of high-quality astronomical imaging data and accurately catalog millions of variable stars through temporal surveying techniques. WFST will conduct a comprehensive and systematic analysis of the Milky Way with a focus on variable stars. As stars whose brightness fluctuates over time, variable stars are invaluable for investigating the internal structure and evolutionary processes of stars. The high sensitivity and multi-band observation capabilities of WFST enable it to precisely detect subtle variations in the brightness of variable stars. Furthermore, WFST excels at detecting faint objects. The high sensitivity and spatial resolution of WFST allow it to have a significant advantage in detecting dim objects and faint signals from low-brightness variables \citep{2023SCPMA..6609512W}. During the commissional phase, WFST J053009.62 + 594557.0 is identified as a new pulsating WD, demonstrating its ability to detect faint and subtle variable stars.

\begin{figure*}{}
	\includegraphics[width=\textwidth, angle=0]{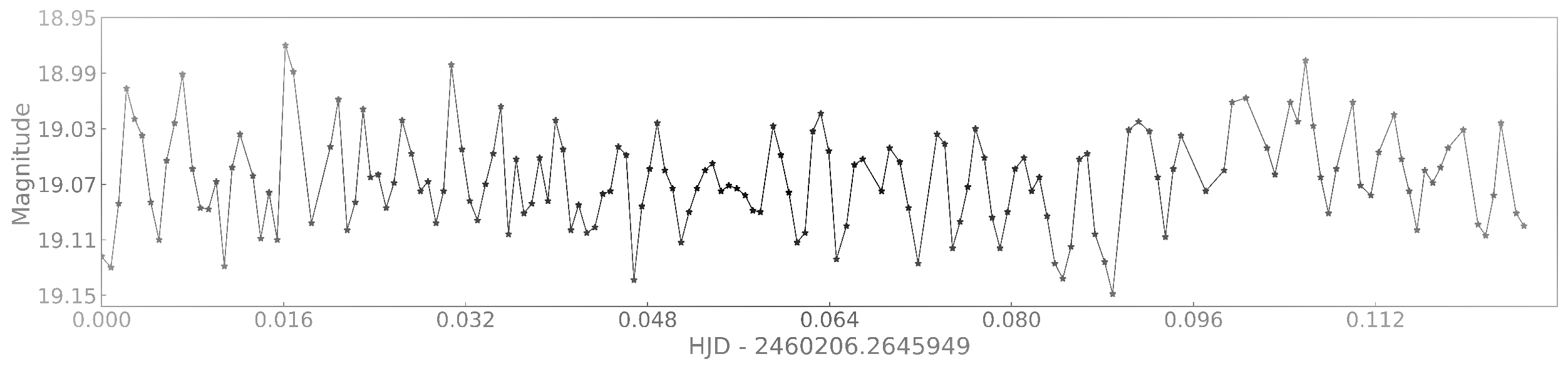}
	\caption{ WFST $g$-band light curve for WFST J0523, obtained on 2023 September 18.}
	\label{fig:lc}
\end{figure*}

\subsection{Photometric Observations}
WFST J0530 was photometrically observed with WFST in $g$ band on 2023 September 18 and November 16. Single exposure was 20\,s with an overhead of approximately 30\,s. As described in \cite{Lin+etal+2025+WFST}, WFST raw images were bias-subtracted and flat-corrected using a modified version of the LSST pipeline \citep{2017ASPC..512..279J,2025arXiv250115018C}. 
Sources exceeding the detection thresholds were extracted from the background-subtracted processed images using point spread function (PSF) modeling.
Astrometric calibration was performed against Gaia DR3 \citep{2016A&A...595A...1G,2023A&A...674A...1G}, and the photometric calibration of the g-band was referenced to Pan-STARRS DR2 \citep{2016arXiv161205560C,2020ApJS..251....6M,2020ApJS..251....7F}.
All fluxes were converted to AB magnitudes. Photometric measurements affected by blending, saturation, or cosmic-ray contamination were discarded. The WFST J0530 light curve was constructed by selecting photometric measurements within a $1^{\prime\prime}$ matching radius. Detailed observation list is provided in Table \ref{tab:obs_logs}, and the partial light curve is presented in Figure \ref{fig:lc}.

\begin{table}
	\centering
	\caption{The list of photometric and spectroscopic observations.}
	\label{tab:obs_logs}
	\begin{tabular}{lccr} 
		\hline
		Telescope        &Instrument &UTC date   &Exp. (s)      \\
		\hline
		WFST         &9k$\times$9k CCD\,     &2023-09-18\,      &20\,     \\
                     & $g$ band      &2023-11-16        &20       \\
		P200         &DoubleSpec\,     &2024-02-01\,      &900\,         \\
		\hline
	\end{tabular}
\end{table}

\subsection{Spectroscopic Observations}

\begin{figure}
    \centering
	\includegraphics[width=0.8\textwidth, angle=0]{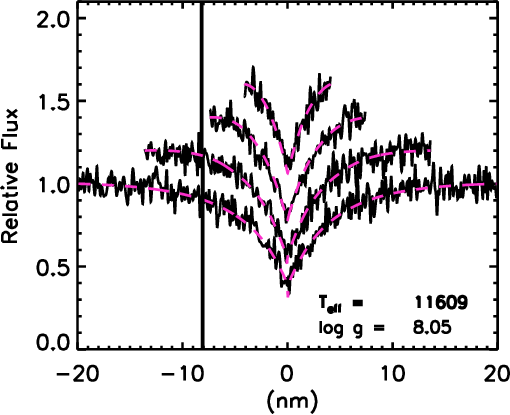}
	\caption{The cut-outs of normalized Balmer lines from P200/DBSP spectrum of WFST J0530. From bottom to top, black lines are the H$\alpha$, H$\beta$, H$\gamma$, and H$\delta$ lines, while the red dashed lines display the best-fitting model.}
	\label{fig:spec}
\end{figure}

Soon after the discovery of WFST J0530, we obtained an optical spectrum using the Double Spectrograph (DBSP) mounted on the 200-inch Hale Telescope at Palomar Observatory \citep[P200,][]{Oke1982}. The observation was carried out with the dichroic D55, which splits the incoming light at 5500~\AA\ into separate blue and red channels. The blue arm used a grism of 600 lines per mm blazed at 3780~\AA, while the red arm employed a grism of 316 lines per mm blazed at 7150~\AA. The spectrum was acquired using a 1.5\arcsec\ slit with an exposure of 1800 seconds. Data reduction was performed using the Python-based \texttt{PypeIt} package \citep{Pypeit1, Pypeit2}, which can implement the standard reduction procedure for long-slit spectra highly automatically. Data are reduced with \texttt{IRAF}, following the standard procedure. Dark and bias subtraction, cosmic-ray removal, and one-dimensional spectral extraction are applied. The spectral parameters were obtained using the spectral fitting method described by \cite{2022MNRAS.509.2674G}, and were ultimately determined to be $T_{\rm eff}$ = 11,609 $\pm$ 605\,K, log\,$g$  = 8.05 $\pm$ 0.36, $M$ = 0.63 $\pm$ 0.22\,$M_{\odot}$. In Fig.\,\ref{fig:spec}, Balmer lines of the observed spectrum are shown, along with the best-fitting spectral models.

\section{Asteroseismological Analysis}
\subsection{Frequency analysis and Mode Identification}

\begin{figure*}
    \centering
	\includegraphics[width=0.8\textwidth]{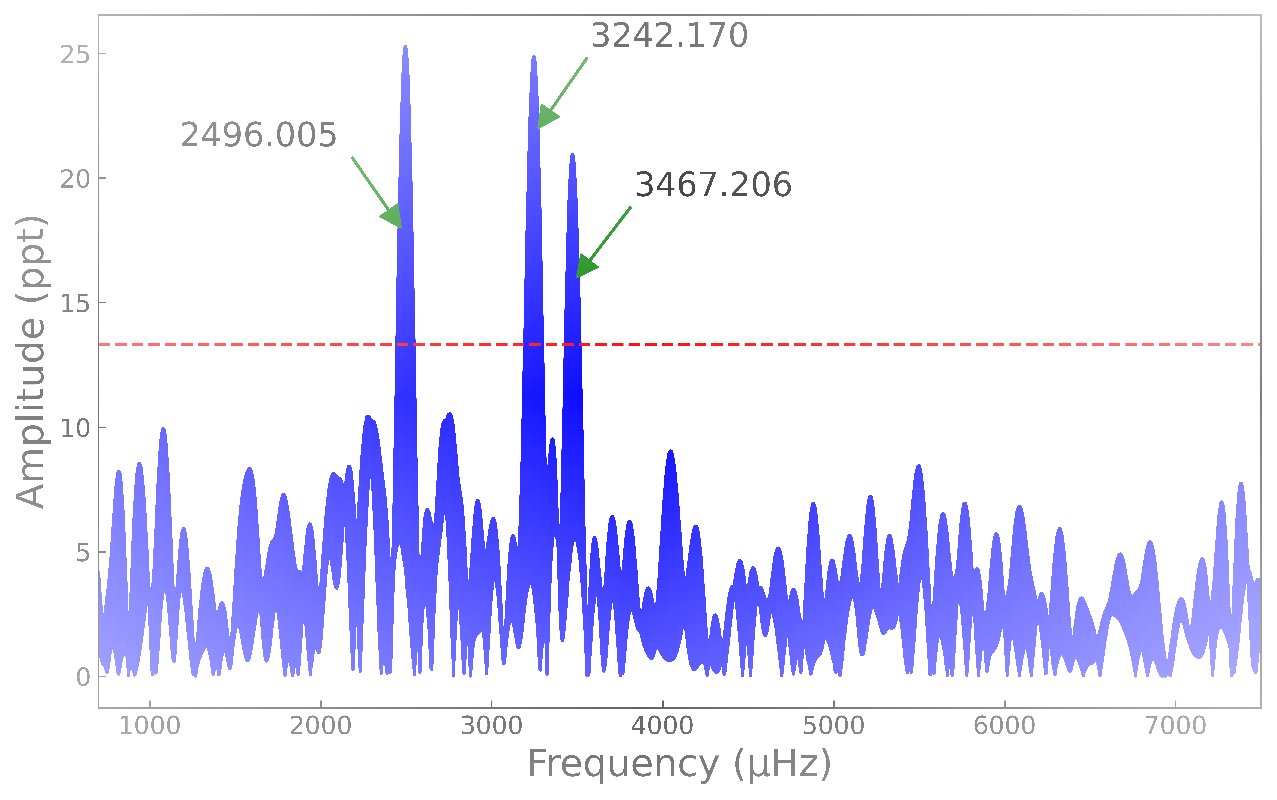}
	\caption{The Lomb–Scargle periodogram of WFST J0530. The horizontal red dashed line indicates the position where \(SNR\) = 4.}
	\label{fig:LS_figure}
\end{figure*}

In this study, a detailed analysis of the light curve was conducted using the \texttt{Period04} \citep{2005CoAst.146...53L}. The light curve was first transformed into the frequency domain via the Fourier transform (FT). Then a standard pre-whitening procedure was applied to the resulting FT spectrum to accurately extract the frequency, amplitude, and phase information of the underlying signal. A signal-to-noise ratio (\(SNR\)) threshold of 4 was set, enabling the identification and extraction of three significant frequencies from the light curve. Figure\,\ref{fig:LS_figure} illustrates the Lomb–Scargle periodogram of these three frequencies, with the red dashed line indicating the threshold \(SNR\)=4. The uncertainties of the frequencies and amplitudes were determined using the Monte Carlo simulation method. Table\,\ref{tab:freq} summarizes the relevant data: the second and third columns list the frequencies and their uncertainties; the fourth and fifth columns present the periods and their uncertainties; the sixth and seventh columns show the amplitudes and their uncertainties; and the final column provides the values of \(SNR\).

The characteristics of non-radial pulsation modes for spherically symmetric stars are described by the spherical harmonic indices \(k\), \(l\), and \(m\) \citep{2008ARA&A..46..157W}. The radial node number is indicated by \(k\) (the radial index), while the absolute value of \(m\) (the azimuthal index) determines the angular distribution of the pulsation modes on the stellar surface, ranging from \(-l\) to +\(l\). When a star rotates, the pulsation-mode degeneracy is lifted, causing the modes to split into 2\(l\)+1 components, where \(l\) is the angular degree of the mode. However, due to geometric cancelation effects \citep{1977AcA....27..203D}, only the \(l\)=1 and \(l\)=2 are predominantly observed. From the light curve of WFST J0530, three independent frequencies were identified, with no evidence of frequency combinations. The absence of observable frequency splitting among these three frequencies suggests that they likely correspond to \(l\)=1 or \(l\)=2 modes, which were assumed in the subsequent asteroseismological analysis.

\begin{table*}
	\centering
	\caption{Frequencies derived from WFST photometric observations. $Freq.$ is frequency in $\mu$Hz, $\delta\,Freq.$ is frequency splitting value in $\mu$Hz, $Per.$ is periods in seconds, and $Amp.$ is amplitude in $ppt$.}
    \label{tab:freq}
	\begin{threeparttable}
		\begin{tabular}{ccccccccccccccccccccccccc}
			\hline
			$ID$	&$Freq.(\mu$Hz)   &$\delta\,Freq.$($\mu$Hz)  & $Per.(s)$  &$\delta\,Per.(s)$   &$Amp.(ppt)$   &$\delta\,Amp.(ppt)$  &$SNR$              \\
			\hline
			$f_{1}$		&\, 2496.005      &\,  0.030     &\, 400.640     &\, 0.005        &\, 24.198    &\,   0.003          &\, 6.33                       \\
			$f_{2}$		&\, 3247.170      &\,  0.008     &\, 308.435     &\, 0.001        &\, 20.737    &\,      0.003       &\, 6.99                       \\
			$f_{3}$		&\, 3467.206      &\, 0.013      &\, 288.417     &\,  0.001       &\, 15.156    &\,       0.003      &\, 5.12                       \\
			\hline
		\end{tabular}
	\end{threeparttable}
	\centering
\end{table*}

\subsection{Theoretical Models}

In the V16 version of \texttt{WDEC} \citep[\texttt{White Dwarf Evolution Code},][]{2018AJ....155..187B}, the \texttt{MESA} \citep[\texttt{Modules for Experiments in Stellar Astrophysics},][] {2018ApJS..234...34P,2019ApJS..243...10P} opacity tables and equation of state are utilized, and 15 input parameters are incorporated. These parameters are specifically included:  effective temperature ($T_{\rm eff}$($K$)), WD mass ($M_{*}$/$M_{\odot}$), total envelope mass (-log($M_{\rm env}/M_{\rm *}$)), helium layer mass (-log($M_{\rm He}/M_{\rm *}$)), hydrogen atmosphere mass (-log($M_{\rm H}/M_{\rm *}$)), helium abundance in the mixed C/He/H region ($X_{\rm He}$), helium diffusion coefficient at the bottom of the envelope, helium diffusion coefficient at the bottom of the pure helium layer, mixing length parameter for the convective zone ($\alpha$), and six oxygen abundance ($X_{O}$) parameters ($h_{1-3}$, $w_{1-3}$, with $h_{1}$ representing oxygen abundance).

In the evolving DAV star models, the convective zone was modeled using standard mixing length theory \citep{1971A&A....12...21B} with a mixing length parameter of $\alpha$ = 0.6 \citep{1995PASP..107.1047B}. The helium diffusion coefficients at the bottom of the envelope and the pure helium layer were fixed at 0.12 \citep{2018AJ....155..187B}. The $M_{*}$/$M_{\odot}$ and $T_{\rm eff}$($K$) parameters were determined based on the well-established properties of DAV stars, ensuring plausible values within reasonable ranges. The -log($M_{\rm env}/M_{\rm *}$) was systematically varied from thick to thin configurations to fully cover all potential variations. The -log($M_{\rm He}/M_{\rm *}$) and the -log($M_{\rm H}/M_{\rm *}$) were set to empirical values of $10^{-2} - 10^{-4}$ and $10^{-4} - 10^{-10}$ \citep{2008PASP..120.1043F,2008MNRAS.385..430C,2009MNRAS.396.1709C,2013ApJ...779...58R,2019MNRAS.490.1803R}, respectively. To account for the possibility of a thinner -log($M_{\rm env}/M_{\rm *}$), the range of helium layer mass was extended to $10^{-5}$. In constructing the evolutionary models, -log($M_{\rm He}/M_{\rm *}$) was maintained less than -log($M_{\rm env}/M_{\rm *}$), and the maximum -log($M_{\rm H}/M_{\rm *}$) was set to $10^{-2}$ times -log($M_{\rm He}/M_{\rm *}$) \citep{1990ApJS...72..335T}. The values of the remaining seven parameters are provided in column 2 of Table\,\ref{tab:model_para}, while their corresponding coarse step ranges are listed in column 3. Based on these specified parameter settings, a total of 14,486,688 models were constructed \citep{ 2024MNRAS.528.5242G}, and asteroseismological analysis of WFST J0530 was performed using these models.

In the fitting process, the coarse steps were first used to preliminarily determine a baseline optimal model, which served as the initial optimal model. Next, centered on the initial optimal model, the grid range was incrementally reduced. Using middle and fine steps, the model underwent refined evolutionary adjustments and parameter tuning. After multiple iterative optimizations, the optimal model was precisely identified, thereby finalizing the fitting process.

\begin{table*}
	\centering
	\caption{Input Model Parameters and Optimal Model of DAV Stars Evolved by \texttt{WDEC}.}
    	\label{tab:model_para}
	\begin{threeparttable}
		\begin{tabular}{cccccccccccccc}
			\hline
			Parameters                   &grid                 &coarse         &middle           &fine        &Model 1      &Model 2 (Optimal model)         \\
			&sizes             &steps        &steps         &steps   &         &               & \\
			\hline
			$M_{*}$/$M_{\odot}$          &\, 0.500 to 0.850        &\, 0.01         &\, 0.005         &\, 0.005     &\,  0.740 $\pm$ 0.005     &\, 0.600 $\pm$ 0.005  \\
			$T_{\rm eff}$($K$)             &\, 10,600 to 12,600      &\, 250          &\, 50            &\, 10         &\,  10,600 $\pm$ 10        &\, 11,850 $\pm$ 10  \\
			-log($M_{\rm env}/M_{\rm *}$)&\, 1.50 to 2.00          &\, 0.50         &\, 0.50          &\, 0.01    &\,  2.00 $\pm$ 0.01         &\, 1.50 $\pm$ 0.01  \\ 
			&\, 2.00 to 3.00          &\, 1.00         &\, 0.50            \\
			-log($M_{\rm He}/M_{\rm *}$) &\, 2.00 to 5.00          &\, 1.00         &\, 0.50          &\, 0.01     &\,  5.00 $\pm$ 0.01         &\, 4.00 $\pm$ 0.01  \\
			-log($M_{\rm H}/M_{\rm *}$)  &\, 4.00 to 10.00         &\, 1.00         &\, 0.50          &\, 0.01     &\,  8.00 $\pm$ 0.01        &\, 7.00 $\pm$ 0.01  \\
			$X_{\rm He}$                 &\, 0.10 to 0.90          &\, 0.16         &\, 0.05          &\, 0.01      &\,  0.26 $\pm$ 0.01       &\, 0.26 $\pm$ 0.01  \\
			\hline
			$X_{O}$ of core center                                                                                           \\
			\hline
			$h_{1}$                         &\, 0.60 to 0.75        &\, 0.03         &\, 0.02          &\, 0.01     &\,  0.65 $\pm$ 0.01         &\, 0.61 $\pm$ 0.01  \\
			$h_{2}$                         &\, 0.64 to 0.71        &\, 0.03         &\, 0.02          &\, 0.01      &\,  0.64 $\pm$ 0.01             &\, 0.64 $\pm$ 0.01  \\
			$h_{3}$                         &\, 0.85                &                &\, 0.02          &\, 0.01              &\,  0.84 $\pm$ 0.01              &\, 0.86 $\pm$ 0.01  \\
			$w_{1}$                         &\, 0.32 to 0.38        &\, 0.03         &\, 0.02          &\, 0.01     &\,  0.38 $\pm$ 0.01            &\, 0.33 $\pm$ 0.01  \\
			$w_{2}$                         &\, 0.42 to 0.48        &\, 0.03         &\, 0.02          &\, 0.01    &\,  0.45 $\pm$ 0.01              &\, 0.45 $\pm$ 0.01  \\
			$w_{3}$                         &\, 0.09                &                &\, 0.02          &\, 0.01              &\,  0.09 $\pm$ 0.01              &\, 0.08 $\pm$ 0.01   \\
			\hline
		\end{tabular}
	\end{threeparttable}
	\centering
\end{table*}

\subsection{Model Matching}

To evaluate the quality of the fit, the average fitting error between the observed and theoretical oscillation periods can be calculated using a merit function. This fitting can be used to compare the quality of different models. 
\begin{equation}
	\chi^{2}=\frac{1}{N} \sum_{i=1}^{N}(P_{\rm obs}-P_{\rm cal})^2,
	\label{eq:quadratic}
\end{equation}
In the above equation, $N$ represents the total number of periods to be fitted, $P_{obs}$ denotes the observed period, and $P_{cal}$ indicates the theoretical oscillation period. During the fitting process for WFST J0530, three observation periods were used to constrain the theoretical oscillation periods: 288.417 s, 308.435 s and 400.640 s.

\begin{figure}
    \centering
	\includegraphics[width=0.8\textwidth]{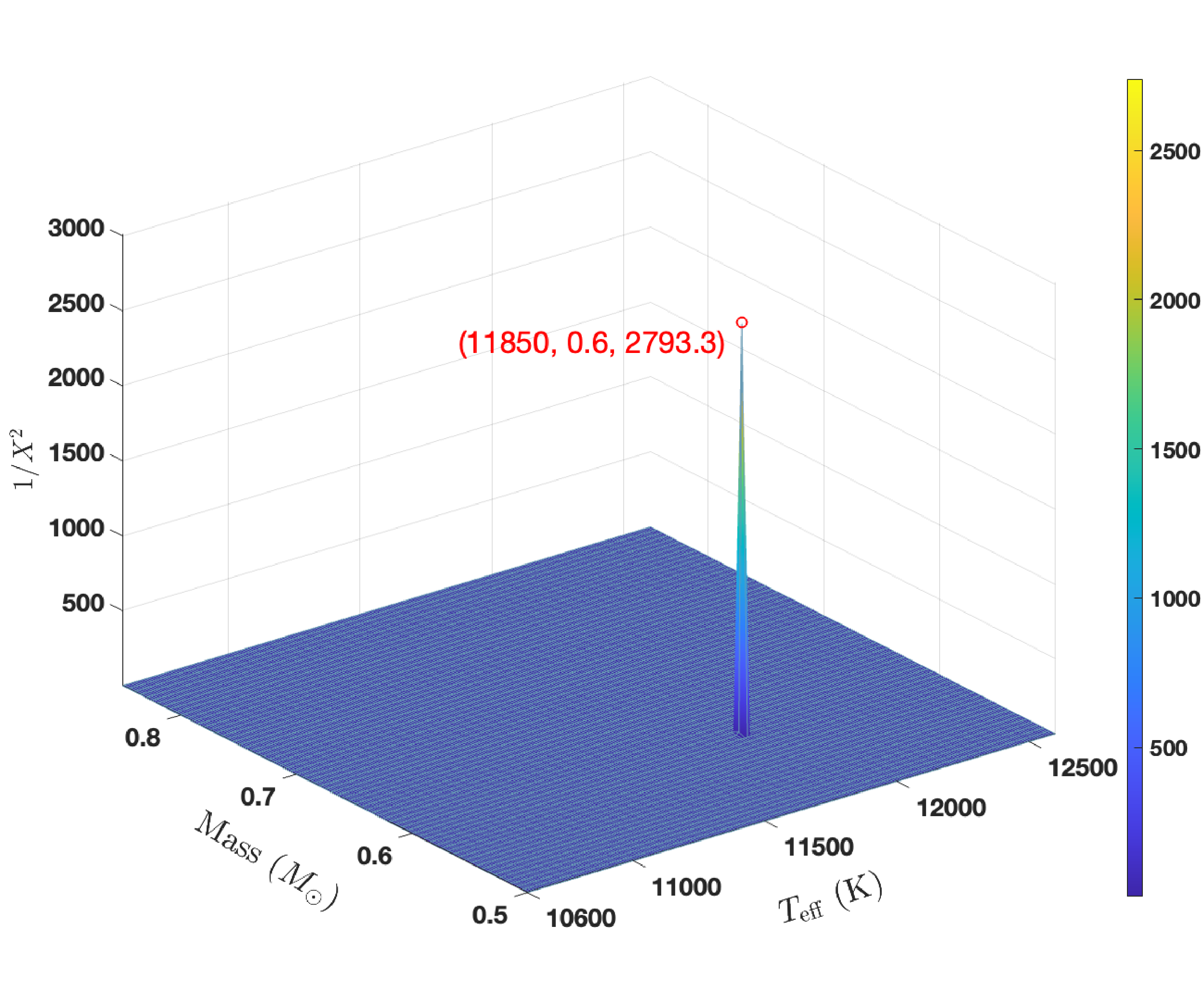}
	\caption{Three-dimensional color graph of function $\chi^2$ versus mass and $T_{\rm eff}$.}
	\label{fig:teff_mass}
\end{figure}

\begin{figure}
    \centering
	\includegraphics[width=0.8\textwidth]{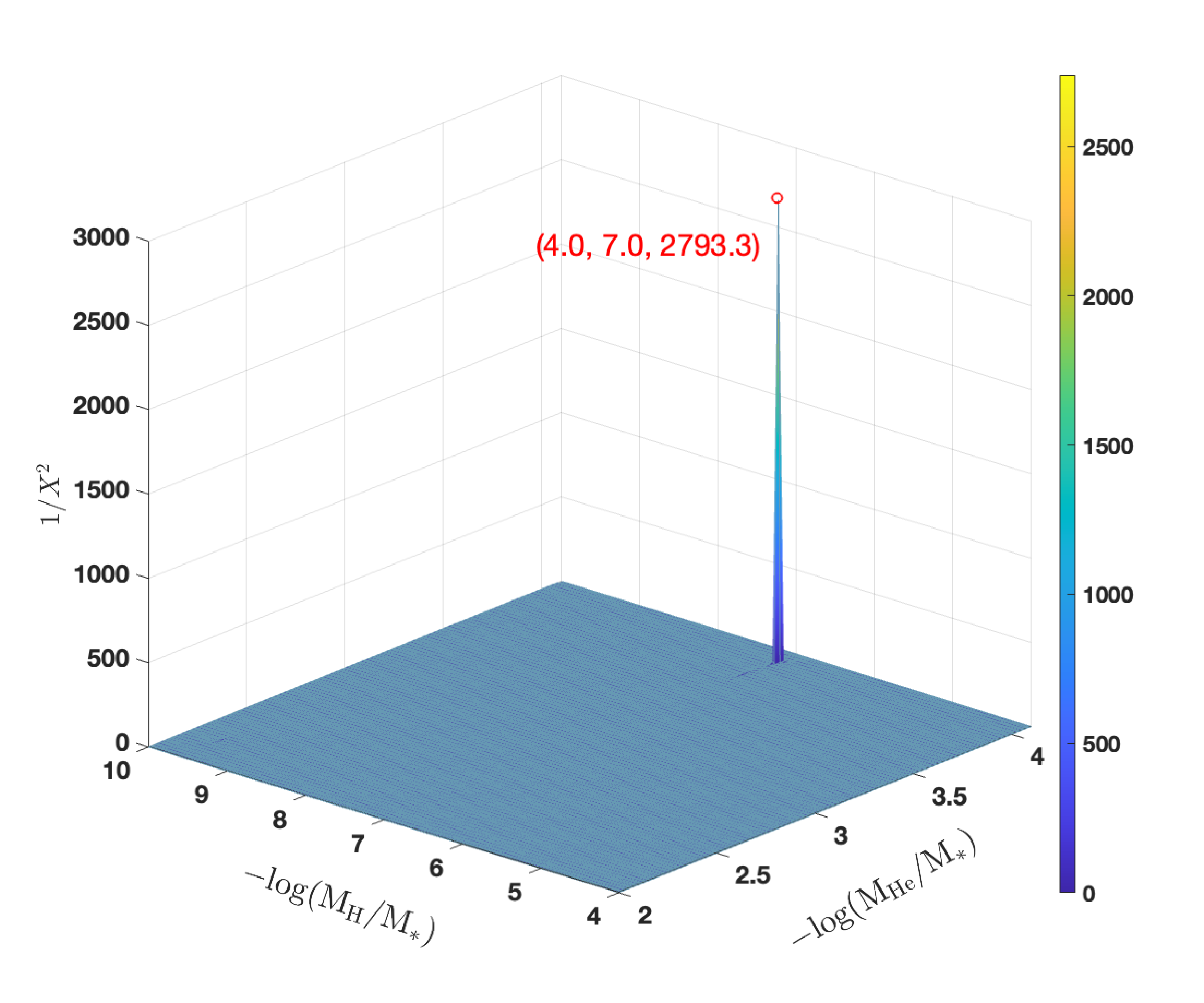}
	\caption{Three-dimensional color graph of function $\chi^2$ versus -log($M_{\rm He}/M_{\rm *}$) and -log($M_{\rm H}/M_{\rm *}$).}
	\label{fig:mhe_mh}
\end{figure}

After sequential fitting through coarse, middle, and fine steps, the least fitting error was found to be $\chi^2$ = $6.2\times10^{-5}$\rm s. The key parameters for the corresponding model, labeled as Model 1, include $T_{\rm eff}$ = 10,600\,K, log\,$g$ = 8.25, and $M_{*}$/$M_{\odot}$  = 0.74. However, Model 1 results showed significant deviations from the results of our spectral fitting (see Table\,\ref{tab:diff_results}). Its $T_{\rm eff}$ is lower than the observed spectroscopic value, while both its mass and log\,$g$ are higher than the observed spectroscopic values. Furthermore, the asteroseismological distance derived from Model 1 exhibited a 27.64\% error relative to the Gaia trigonometric parallax distance. For these reasons, Model 1 has been discarded. Considering only three periods are detected, we extended the searching by including many more matching models than usual. After an extensive searching, the optimal model is found and labeled as model 2 in Table\,\ref{tab:model_para}.

In contrast, Model 2 demonstrated much closer alignment with the spectroscopic observations, whose parameters are $T_{\rm eff}$ = 11,850\,K, log\,$g$ = 8.02, and $M_{*}$/$M_{\odot}$ = 0.60. Moreover, the asteroseismological distance of Model 2 showed a difference of only 1.45\% relative to the trigonometric parallax distance from Gaia DR\,3. Therefore, Model 2 was selected as the final optimal model. All subsequent references to the optimal model in this discussion refer to Model 2.

The fitting error of the optimal model was quantified as $\chi^2$ = $3.58\times10^{-4}$\rm s. The last column of Table\,\ref{tab:model_para} presents the optimal parameters, with their uncertainties following the corresponding parameters. These uncertainties are determined by the full width at half-maximum of $1/\chi^2$. The optimal parameters indicate that WFST J0530 is an intermediate-mass WD with a thin helium layer and a thin hydrogen atmosphere.

\begin{figure*}{}
    \centering
	\includegraphics[width=\textwidth, angle=0]{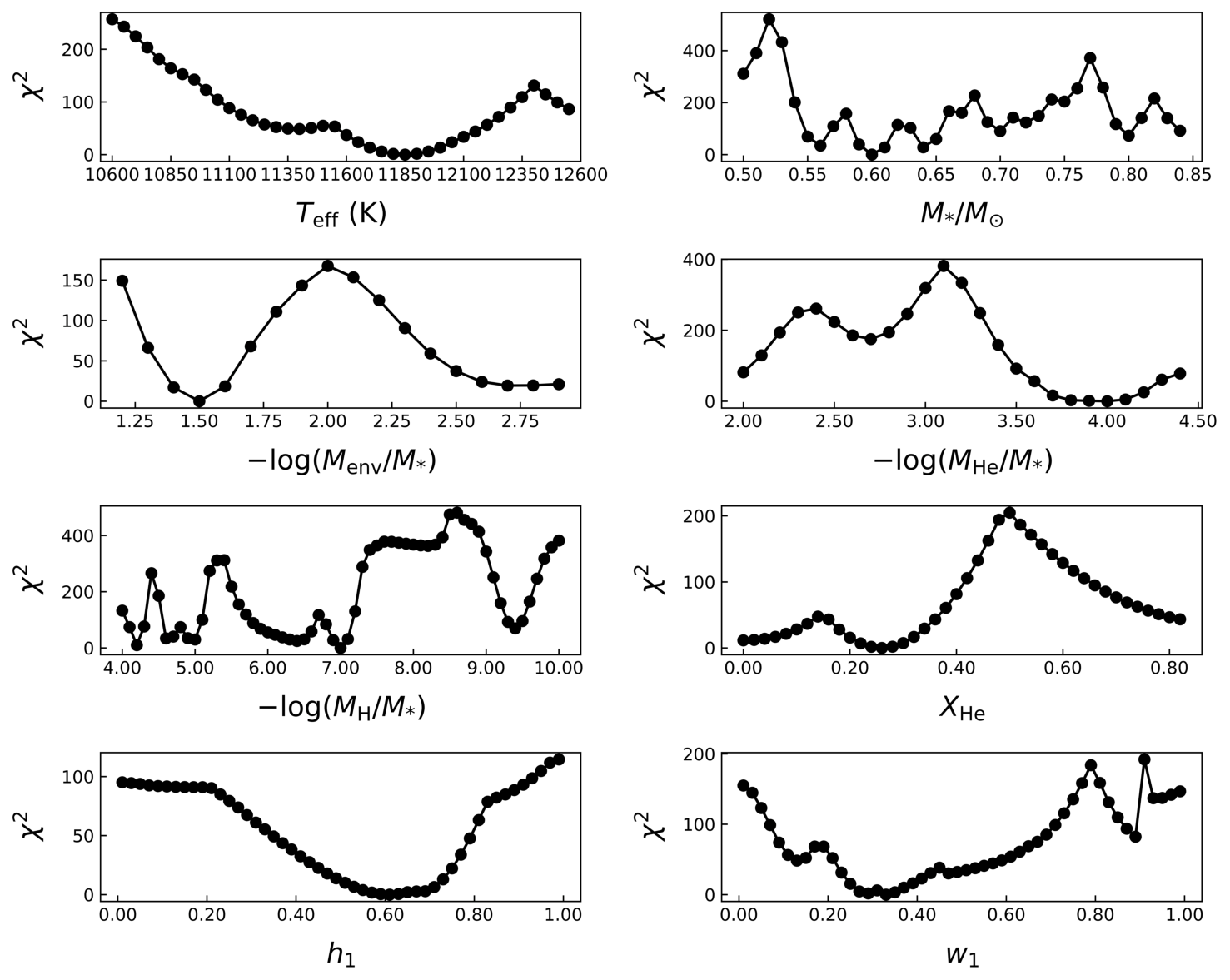}
	\caption{Sensitivity graphs of function $\chi^2$ with respect to optimal parameters $T_{\rm eff}$, $M_{*}$/$M_{\odot}$, -log($M_{\rm env}/M_{\rm *}$), -log($M_{\rm He}/M_{\rm *}$), -log($M_{\rm H}/M_{\rm *}$), $X_{\rm He}$, $h_{1}$, and $w_{1}$ .}
	\label{fig:sensitivity}
\end{figure*}

Figure\,\ref{fig:teff_mass} presents a three-dimensional color graph of the function $\chi^2$ versus $M$ and $T_{\rm eff}$. The horizontal axis shows the $T_{\rm eff}$ of WDs, which range from 10,600\,K to 12,600\,K in step of 10\,K. The y axis represents the $M_{*}$/$M_{\odot}$, ranging from 0.5 $M_{\odot}$ to 0.85 $M_{\odot}$ with step of 0.005 $M_{\odot}$. The Z axis is represented by the reciprocal of $\chi^2$, with a color bar illustrating how the Z axis values change in response to variations in the horizontal and longitudinal coordinates. The optimal model, with effective temperature of 11,850\,K and mass of 0.60 $M_{\odot}$, corresponds to the highest point on the Z-axis. In Figure\,\ref{fig:mhe_mh}, a three-dimensional color graph of the function $\chi^2$ versus -log($M_{\rm He}/M_{\rm *}$) and -log($M_{\rm H}/M_{\rm *}$) is constructed in a similar way. The horizontal axis represents the -log($M_{\rm He}/M_{\rm *}$), plotted from $10^{-2}$ to $10^{-4.1}$ with a step of $10^{-0.02}$. The y axis represents the -log($M_{\rm H}/M_{\rm *}$), plotted in the range of $10^{-4}$ to $10^{-10}$ with a step of $10^{-0.02}$. The highest point on the Z-axis reflects the optimal values of -log($M_{\rm He}/M_{\rm *}$) and -log($M_{\rm H}/M_{\rm *}$), which are $10^{-4}$ and $10^{-7}$, respectively. 
The optimal thickness of -log($M_{\rm He}/M_{\rm *}$) is relatively thin compared to the typical value. The optimal thickness of the -log($M_{\rm H}/M_{\rm *}$) has been determined to be relatively thin compared to the average value of $10^{-6.5}$ \citep{2008MNRAS.385..430C}.

Figure\,\ref{fig:sensitivity} presents a graph of the sensitivity analysis for the parameters listed in the first column of Table\,\ref{tab:model_para}, with the aim of better illustrating how the variations of parameters influence the merit function $\chi^2$. This sensitivity graph highlights the critical parameters whose variations most strongly affect the merit function $\chi^2$, which in turn helps to identify the optimal parameters. Each subplot in Figure\,\ref{fig:sensitivity} represents the sensitivity profile of a different parameter ($T_{\rm eff}$, $M_{*}$/$M_{\odot}$, -log($M_{\rm env}/M_{\rm *}$), -log($M_{\rm He}/M_{\rm *}$), -log($M_{\rm H}/M_{\rm *}$), $X_{\rm He}$, $h_{1}$, and $w_{1}$), with other parameters maintained at their optimal values (as listed in the last column of Table\,\ref{tab:model_para}) while varying the parameter on the horizontal axis. Different $\chi^2$ values can be generated through these parameter variations. The model parameters associated with the minimal $\chi^2$ value are identified as optimal parameters. Furthermore, the sensitivity graph can help researchers identify optimal parameter ranges more efficiently. By tracking how $\chi^2$ varies with parameter changes, researchers can determine the optimal parameter combination for $\chi^2$.

\begin{figure}
    \centering
	\includegraphics[width=0.6\textwidth]{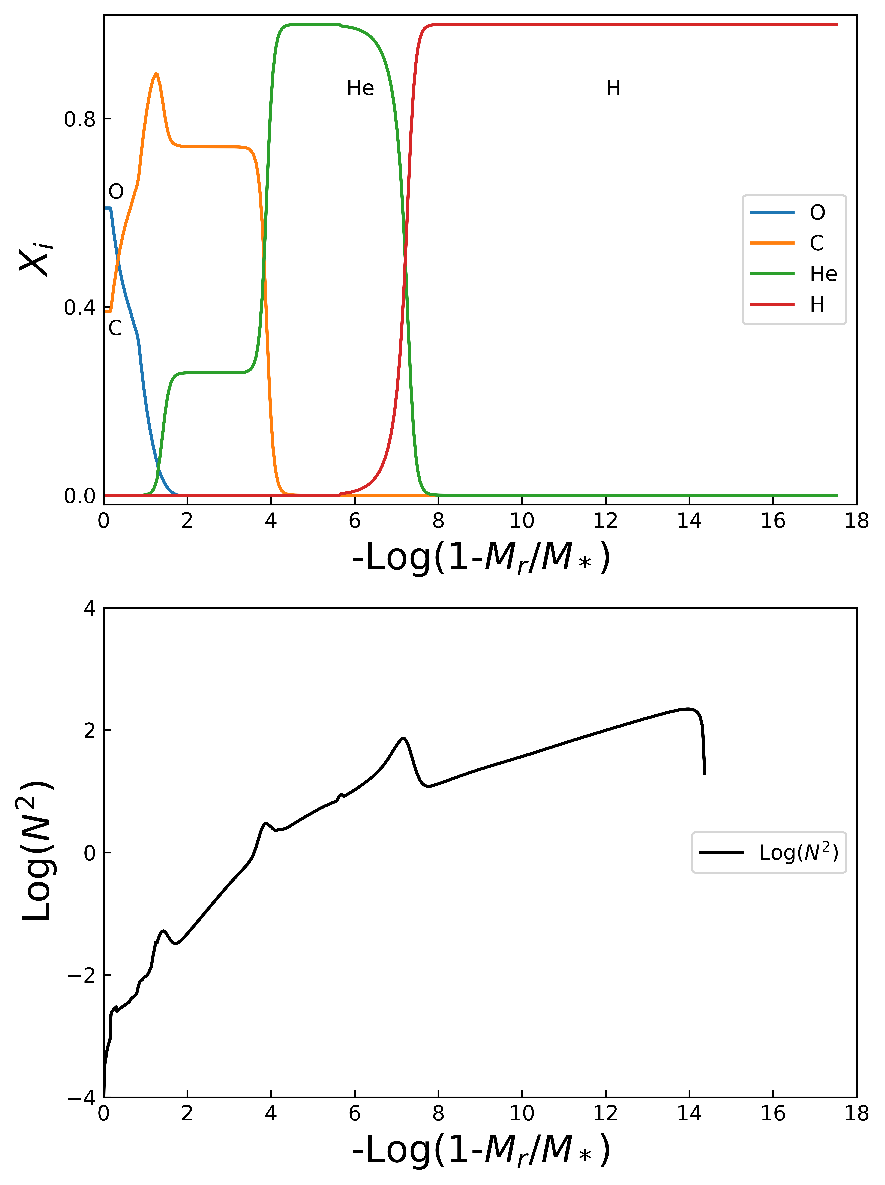}
	\caption{Figure of chemical composition profile and brunt$-$V\"{a}is\"{a}l\"{a} frequency of WFST J0530}.
	\label{fig:Buoyancy_figure}
\end{figure}

Figure\,\ref{fig:Buoyancy_figure} illustrates the chemical composition profile and Brunt$-$V\"{a}is\"{a}l\"{a} frequency of the optimal model. The upper part of the figure shows the relationship between the relative mass $M_{\rm r}/M_{\rm *}$ and the abundance variations of elements O, C, He, and H, extending from the C/O core of the WD outward to the surface hydrogen atmosphere. C/O core of the model originates from the helium nuclear burning of the progenitor star. The adjacent outer region, composed of O, C, and He, is attributed to nucleosynthesis during the TP-AGB phase. The outer He layer primarily forms because of gravitational settling of the elements during H burning. The outermost hydrogen atmosphere comprises primordial hydrogen, deposited before nuclear fusion and influenced by gravitational settling \citep{2019A&ARv..27....7C}. The features of the period spectrum in the g mode are predominantly shaped by the Brunt$-$V\"{a}is\"{a}l\"{a} frequency shown in the lower part of Figure \ref{fig:Buoyancy_figure}. In the central region dominated by the electron degeneracy pressure, the Brunt$-$V\"{a}is\"{a}l\"{a} frequency is very small. As regions move away from the center, the Brunt$-$V\"{a}is\"{a}l\"{a} frequency gradually increases. Specifically, each bump in the Brunt$-$V\"{a}is\"{a}l\"{a} frequency profile corresponds to the gradient transition regions in the chemical composition profile, reflecting its influence on the WD's pulsation on a macroscopic scale.

In Table\,\ref{tab:optimal_periods}, the first column presents the observational periods used to constrain the theoretical model, the second column shows the theoretical oscillation periods of the optimal model, and the third column lists their corresponding errors. The minimal difference between the observed periods and the theoretical oscillation periods indicates a strong correspondence between observation and theory. The results of theoretical calculations demonstrate that two of the three observed periods correspond to the mode \(l\) = 1, while the remaining one corresponds to the mode \(l\) = 2, as indicated in column 4.

\begin{table}
	\centering
	\caption{The table of theoretical oscillation periods for the optimal model. $P_{obs}$ is the observed periods, and $P_{cal}$ is the calculated periods.}
	\label{tab:optimal_periods}
	\begin{tabular}{lccccccccr} 
		\hline
		$P_{obs}$        &$P_{cal}$  &$|P_{obs}-P_{cal}|$   &$l$  &$k$   \\
		(s)              &(s)                    &(s)                                                 \\
		\hline
		288.417         &\,288.436     &\, 0.019   &\, 1 &\, 3     \\
		308.435         &\,308.432     &\, 0.003   &\, 2 &\, 7     \\
		400.640         &\,400.642    &\, 0.002    &\, 1 &\, 5       \\
		\hline
	\end{tabular}
\end{table}

\begin{table*}
	\centering
	\caption{Main parameters of WFST J0530 obtained from various methods. ID1 denotes the spectroscopic fitting results of this study. ID2 signifies the results from Gaia XP spectrum. ID3 corresponds to the Gaia color-magnitude diagram results with extinction. Lastly, ID4 indicates the asteroseismology fitting results of this study.}
	\label{tab:diff_results}
	\begin{threeparttable}	
	\begin{tabular}{lcccccr} 
		\hline
		$ID$   &$M$($M_{\odot}$)   &$T_{\rm eff}$\,(K)                 &log\,$g$             &Source of Data  \\
		\hline
		 1     &\, 0.63 $\pm$ 0.22               &\, 11,609  $\pm$ 605                     &\, 8.05 $\pm$ 0.36                 &\,  Spectral Fitting (this work)  \\
		 2     &\,  0.619$\pm$0.070              &\,  11,657 $\pm$ 505                &\,   8.025  $\pm$ 0.086        &\,    Gaia XP spectrum (Vincent et al. 2024)            \\
		3    &\, 0.63 $\pm$ 0.08              &\, 12,110 $\pm$ 770                        &\,  8.05 $\pm$ 0.13                                                              &\,   Gaia color-magnitude diagram (this work)    \\
		4   &\, 0.600  $\pm$ 0.005                &\, 11,850  $\pm$ 10                 &\,  8.02 $\pm$ 0.01                       &\,  asteroseismology (this work)                      \\			
		\hline
	\end{tabular}
	\end{threeparttable}
	\centering	
\end{table*}

\section{Discussion}
\subsection{Validation from various methods}
Considering only three periods are detected in our observations, various independent methods have been tried to ensure our asteroseismology result is correct. Table\,\ref{tab:diff_results} presents the results of the WD parameters obtained by different methods, including $M_{*}$/$M_{\odot}$, $T_{\rm eff}$ and log\,$g$, along with the data sources. Row 1 presents the spectroscopic results of this study, while row 2 displays the results from Gaia XP spectrum obtained by \cite{2024A&A...682A...5V}. Row 3 shows the Gaia color-magnitude diagram results, which incorporate extinction corrections. The results of the asteroseismological fitting are provided in row 4. 

\subsubsection{Follow-up DBSP spectrum}
As described in Section 2.3, our follow-up spectrum obtained from DBSP at P200 is used for spectral fitting. Following the procedures in \cite{2022MNRAS.509.2674G}, the parameters are derived as $T_{\rm eff}$ = 11,609 $\pm$ 605\,K, log\,$g$  = 8.05 $\pm$ 0.36, M = 0.63 $\pm$ 0.22\,M$_{\odot}$. Within the uncertainties, the spectral results align with our asteroseimology results.

\subsubsection{Gaia parallax}
The luminosity of the theoretical optimal model can be used to estimate the asteroseismology distance \citep{2019A&A...632A..42B,2021tsc2.confE..46U,2024A&A...691A.194C}. Comparing this asteroseismic distance with the Gaia trigonometric parallax distance can effectively verify the accuracy of the optimal model. The luminosity of the optimal model for WFST J0530 has been determined as log($L/L_{\odot}$) = - 2.563. The solar absolute bolometric magnitude, denoted as $M_{bol,\odot}$ , is obtained from \cite{2000asqu.book..499C} and is determined to be 4.74. Based on log($L/L_{\odot}$) and $M_{bol,\odot}$ , the result of the absolute bolometric magnitude ($M_{bol}$) can be calculated as 11.15 mag using the formula $M_{bol} = M_{bol,\odot} - 2.5log(L/L_{\odot})$ \citep{2019A&A...632A..42B}. To convert the absolute bolometric magnitude to the absolute visual magnitude ($M_{V}$), a bolometric correction \citep{1995PASP..107.1047B} of - 0.59 mag is introduced. Then, using the formula $M_{V} = M_{bol}-BC$, the value of $M_{V}$ is calculated to be 11.74 mag. The asteroseismological distance is calculated as 281.84 $\pm$ 1.88 pc using the distance formula 5log$d$ = $m_{V}$ + 5 $-$ $M_{V}$, given the apparent visual magnitude $m_{V}$ = 18.99 mag \citep{2008AJ....136..735L}. In particular, the Gaia distance \citep{2020yCat.1350....0G} for WFST J0530 is 285.94 $\pm$ 18.66 pc. The results obtained from the two different distance measurement methods show an inconsistency of 1.45\%. When uncertainty is taken into account, the two measurements are highly consistent within the error range.

\subsubsection{Gaia XP spectrum}
Based on Gaia XP spectra, \cite{2024A&A...682A...5V} classified and estimated basic parameters of 100,000 high-quality WDs from Gaia, using machine learning techniques and an automated spectral fitting program. From that work, $T_{\rm eff}$, log\,$g$, mass and radii of WFST J0530 are derived as: 11,657 $\pm$ 505\,K, 8.025 $\pm$ 0.086 and 0.619 $\pm$ 0.070\,$M_{\odot}$, respectively. It is also in agreement with our asteroseismology result considering their corresponding errors.

\begin{figure}
    \centering
	\includegraphics[width=0.8\textwidth]{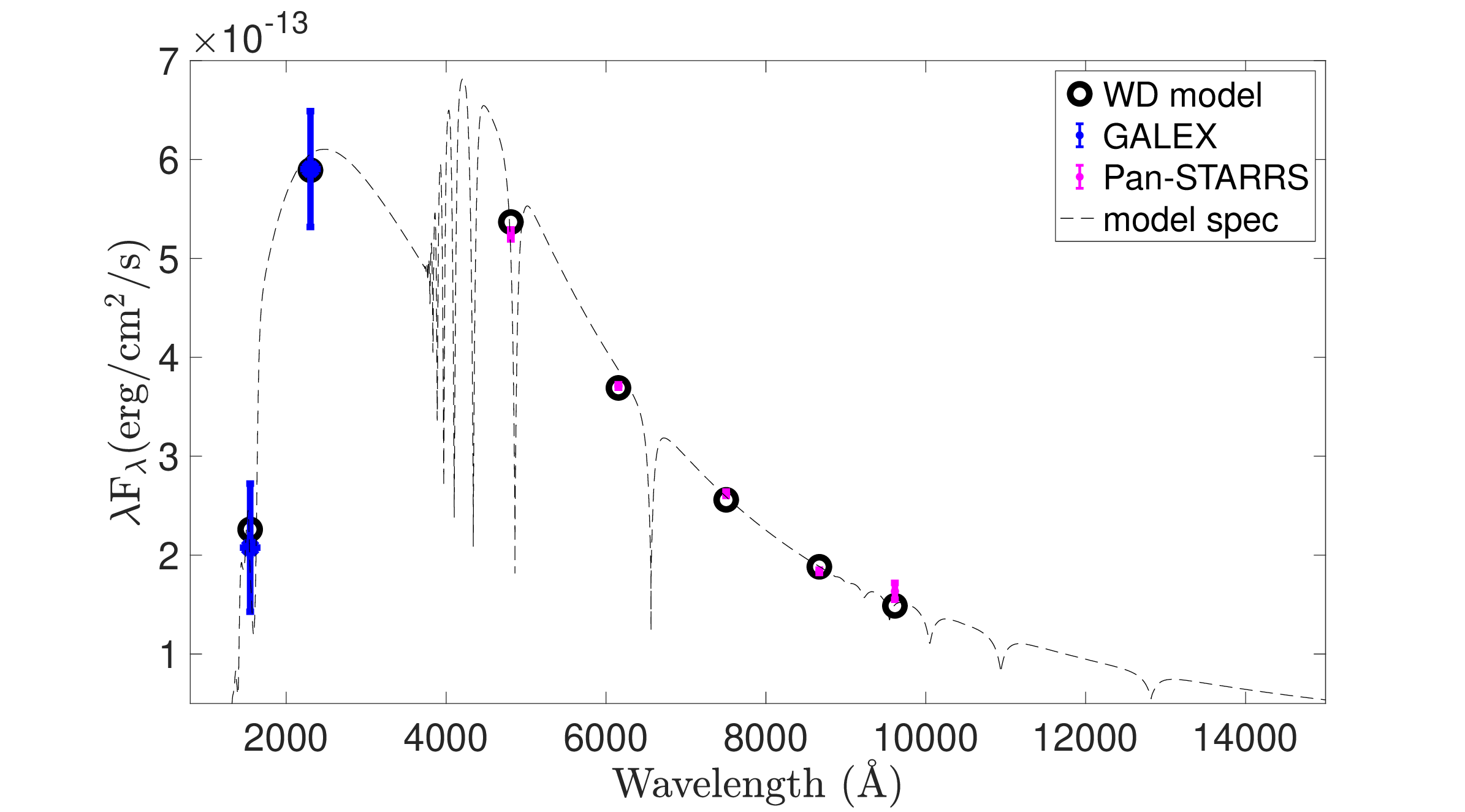}
	\caption{The SED fitting of WFST J0530. Blue and magenta symbols are observed points from GALEX and Pan-STARRS, respectively. Black circles are best fit WD model synthetic photometric points. Dashed line represent the best fit DA WD model spectrum, corresponding to $T_{\rm eff}$=11,000\,K and log\,$g$=8.0.}
	\label{fig:sed}
\end{figure}

\subsubsection{Gaia color-magnitude diagram}
Based on accurate photometric observation and the parallax from Gaia, the location of WFST J0530 can be calculated in the Gaia color-magnitude diagram. With the help of WD evolution models, the $T_{\rm eff}$ and mass of WFST J0530 can be derived. Firstly, according to \cite{2020AJ....159...84B}, E(BP-RP) = 0.0558, the dereddened BP-RP can be calculated as 0.0269 - 0.0558 = -0.0289. Extinction in G band: A$_{G}$ = 1.890*0.0558 = 0.105 \citep{2019gaia.confE..59W}. Absolute G-magnitude: M$_{G}$ = m$_{G}$ - 5*log$d$ + 5 - A$_{G}$ = 11.853 - 0.105 = 11.748. Using the WD model from \cite{2020ApJ...901...93B}, the $T_{\rm eff}$ and the mass are derived to be 12,110 $\pm$ 770\,K and 0.63 $\pm$ 0.08\,$M_{\odot}$, respectively. Within the uncertainties, $T_{\rm eff}$ and mass are consistent with our asteroseismology results.

\begin{figure}
    \centering
	\includegraphics[width=0.8\textwidth]{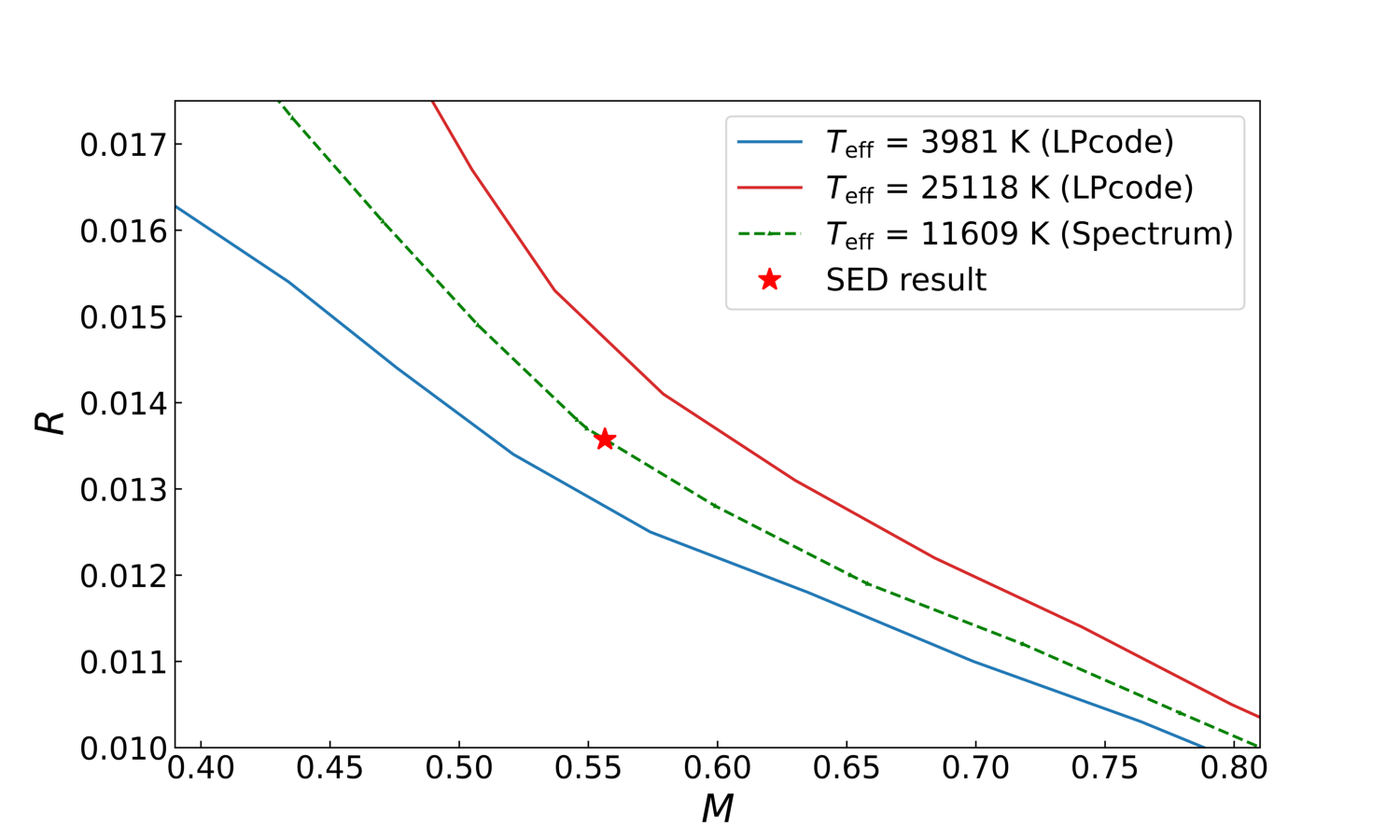}
	\caption{Mass–radius relation for DA WDs. Blue, red, and green dashed curves depict the relations for $T_{\rm eff}$ = 3981\,K, 25,118\,K, and 11,609\,K, respectively. The red star symbol marks the location of WFST J0530.}
	\label{fig:MR}
\end{figure}

\subsubsection{SED fitting}
By collecting photometric data from GALEX and Pan-STARRS, we constructed SED of WFST J0530 (shown in Fig. \ref{fig:sed}). Then in order to obtain the radius of WFST J0530, the SED is used to fit the DA WD model of \cite{Koester2010}. By fixing log\,$g$ to 8.0, the $T_{\rm eff}$ and the proportionality factor M$_{d}=(R/D)^2$ of the best fit model are obtained to be 11,000\,K and $1.118\times10^{-24}$, respectively. The radius of WFST J0530 is derived to be 0.0136\,$R_{\odot}$ or 1.48\,$R_{\oplus}$ through M$_{d}$ and Gaia parallax. By adopting the $T_{\rm eff}$=11,609\,K from spectral fitting, radius of 0.0136\,$R_{\odot}$ and the theoretical mass-radius relation from \textit{LPcode} \citep[Figure\,\ref{fig:MR};][]{2016ApJ...823..158C}, the mass can be derived to be 0.56\,$M_{\odot}$, which is only 0.04\,$M_{\odot}$ lighter than our asteroseismology mass.

\begin{figure}
    \centering
	\includegraphics[width=0.8\textwidth]{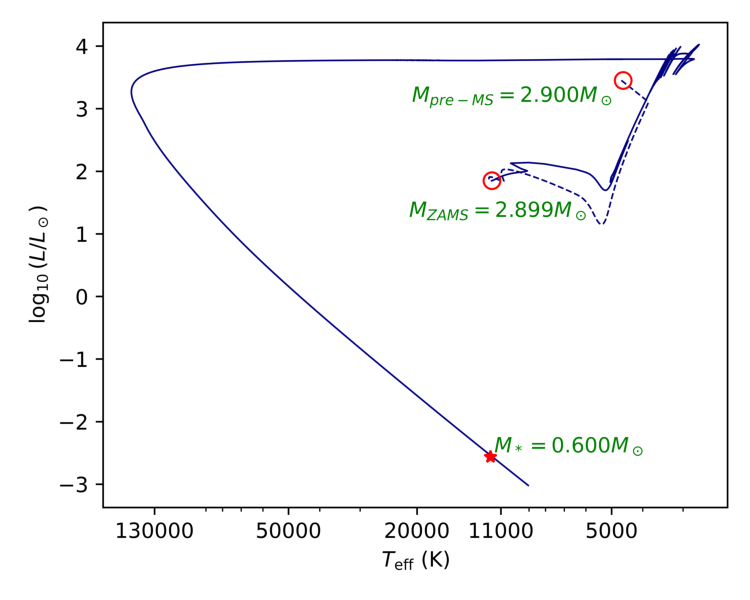}
	\caption{The complete evolutionary trajectory of a pre-main sequence star with an initial mass of 2.9 $M_{\odot}$ evolving to a WD with a final mass of 0.6 $M_{\odot}$. The position of WFST J0530 is marked with a red pentagram. Red circle locates the ZAMS stage with mass of 2.899\,$M_{\odot}$. Dashed line tracks the pre-main sequence phase evolution.}
	\label{fig:evo_track}
\end{figure}

\subsection{Evolutionary history of WFST J0530}

Figure\,\ref{fig:evo_track} displays the complete evolutionary track of a 0.6 $M_{\odot}$ WD evolved from a 2.9 $M_{\odot}$ progenitor, as simulated with \texttt{MESA} release 8118 (WDEC V16 based on). During the evolutionary process, we adjusted only the initial mass to 2.9 $M_{\odot}$ and set the luminosity termination condition to log($L/L_{\odot}$) = -3. All other parameters were kept at their default settings in the \texttt{$1$\(M\_pre\_ms\_to\_wd\)} module. In the figure, the blue dashed line illustrates the evolution of the star from the pre-main-sequence phase to the zero-age main sequence (ZAMS). Upon reaching the ZAMS, the star has a mass of 2.899 $M_{\odot}$, a radius of 2.0 $R_{\odot}$, and a luminosity of log($L/L_{\odot}$) = 1.8. The blue solid line depicts the star's evolution from the ZAMS to the WD stage. The red pentagram indicates current position of the WFST J0530.

\subsection{Comparison with other DAVs}

\begin{figure}
    \centering
	\includegraphics[width=1.0\textwidth]{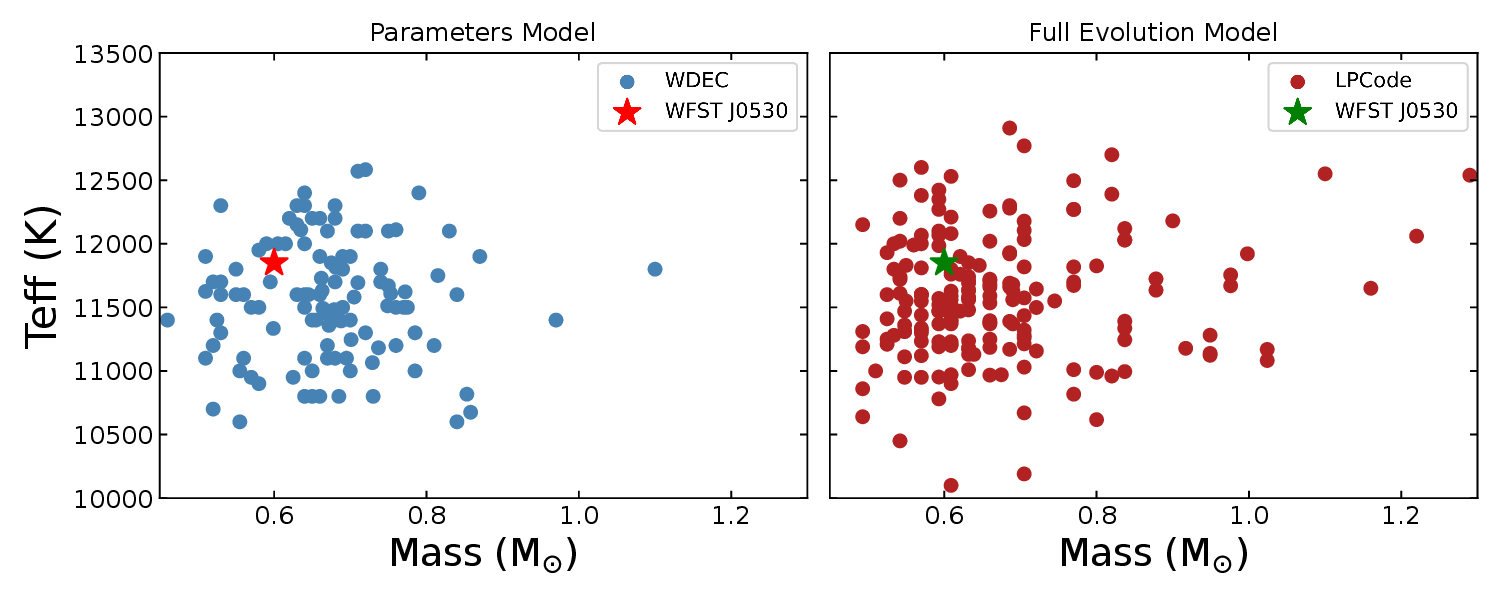}
	\caption{Asteroseismological result comparison of $T_{\rm eff}$ versus WD mass between parameterized and full evolutionary models for DAV stars. The location of WFST J0530 is marked with a five-pointed star.}
	\label{fig:mass_teff_comp}
\end{figure}

\begin{figure}
    \centering
	\includegraphics[width=1.0\textwidth]{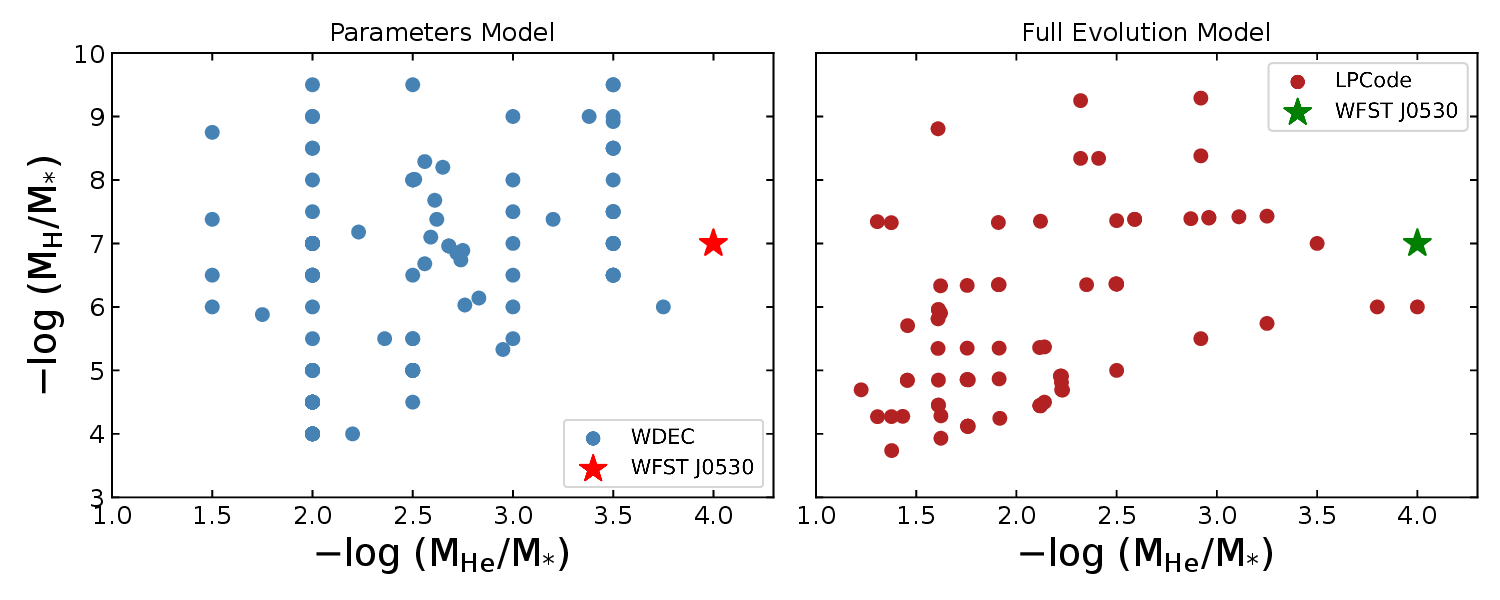}
	\caption{Comparison of helium layer masses and hydrogen atmosphere masses from asteroseismological results of DAVs derived using parameterized and full evolutionary models. The location of WFST J0530 is marked with a five-pointed star.}
	\label{fig:mhe_mh_comp}
\end{figure}

\begin{figure}
    \centering
	\includegraphics[width=1.0\textwidth]{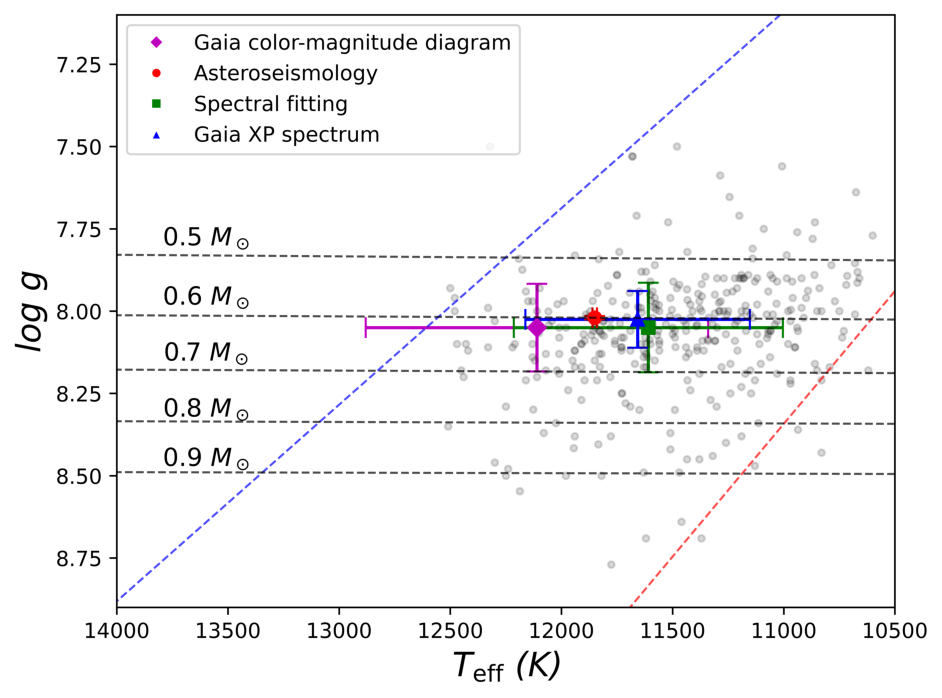}
	\caption{The result locations of WFST J0530 derived from different methods in the $T_{\rm eff}$ and log\,$g$ space with pulsational instability strip. The theoretical blue and red edges are from \protect\cite{2015ApJ...809..148T}. The results obtained from asteroseismological analysis, spectral fitting, Gaia XP spectrum, and Gaia color-magnitude diagram are marked with red circles, green squares, blue triangles, and magenta diamonds, respectively.}
	\label{fig:results_comp}
\end{figure}

In the optimal model of WFST J0530, the oxygen abundance is 0.61, consistent with the results ($X_{o}$ = 0.5–0.8) derived by \cite{2010ApJ...716.1241S} based on stellar structure and evolution. Some parameterized asteroseismological studies have also reported high oxygen abundances ($X_{o}$ $>$ 0.8), as seen in the works of \cite{2018Natur.554...73G}, \cite{2021arXiv210703797C}, \cite{2021A&A...651A..14B}, \cite{2022ApJ...934...34C}, and \cite{2022FrASS...9.9045G}. One of the characteristics of the chemical composition profile of this work is the presence of a pure carbon buffer at -$log(1 - Mr/M*)$ = 1.6 $-$ 3.5. The similar structure was first discovered in the asteroseismological model of the DBV star KIC 08626021 \citep[Figure 2 in ][]{2018Natur.554...73G}, and has since appeared only in subsequent parameterized studies (\citep{2019A&A...632A..42B, 2022FrASS...9.9045G, 2023MNRAS.523.1591G, 2023ApJ...948...74H, 2023MNRAS.522.6094Y, 2024ApJ...976...46Z}). This structural feature may be a common outcome in parameterized models.

In Figure\,\ref{fig:mass_teff_comp}, we systematically compare the fitting results of parameterized and full evolutionary models for DAVs, with data sourced from references \cite{2008MNRAS.385..430C},\cite{2009MNRAS.396.1709C}, \cite{2012MNRAS.420.1462R}, \cite{2013ApJ...779...58R}, \cite{2021A&A...651A..14B}, \cite{2022MNRAS.511.1574R}, \cite{2023ApJ...948...74H}, \cite{2023A&A...669A..62B}, \cite{2023MNRAS.523.1591G}, \cite{2023MNRAS.518.1448R}, \cite{2024MNRAS.528.5242G}, \cite{2025ApJ...988...32C}, and \cite{2025ApJ...984..112R}. The left panel shows the mass–effective temperature distribution derived from the parametric models, while the right panel corresponds to the full evolutionary models. Both panels show that the sample masses are concentrated in the range 0.5–0.9 $M_{\odot}$ and that the effective temperatures all fall within the pulsation instability strip. Within the mass range exceeding 0.9 $M_{\odot}$, the number of samples covered by the full evolutionary model significantly surpasses those of the parameterized model. This discrepancy can be attributed to the relatively limited application of the WDEC parameterized model in fitting high-mass WD stars. The asteroseismic result of WFST J0530 is marked by a five-pointed star. Compared with other DAVs, the mass and temperature of WFST J0530 are both moderate.

Figure\,\ref{fig:mhe_mh_comp} displays the distributions of helium layer mass and hydrogen atmosphere mass derived from fitting DAVs with different models. The left panel shows the results of the parametric model, while the right panel presents those of the full evolutionary model. Overall, the full evolutionary models tend to concentrate towards the thick helium layers and thick hydrogen atmosphere region. The five-pointed star representing WFST J0530 exhibits an extremely thin helium layer and a moderate hydrogen atmosphere mass.

\section{Conclusion and Summary}

In this paper, we present an asteroseismological analysis of WFST J0530, a newly discovered DA-type pulsating WD. It was identified in WFST and has a {\it Gaia} G-band magnitude of 19.13~mag \citep{2023A&A...674A..33G} or SDSS g-band magnitude of 19.11\,mag, thus placing it among the fainter known DAV stars. For comparison, the previously identified faintest DAV star is SDSS J121606.32+092345.1, which has a g-band magnitude of 21.76 \citep{2005A&A...442..629K, 2025arXiv250803184C}. This study analyzed two nights of photometric data from WFST. Three period signals were successfully extracted using the \texttt{Period04}. In the asteroseismological analysis, we constructed a parameterized DA-type pulsating WD model using the \texttt{WDEC} V16. Employing the three observed periods as constraints, we fitted the model and obtained the optimal parameters: $T_{\rm eff}$ = 11,850 $\pm$ 10\,K, log\,$g$ = 8.02 $\pm$ 0.01, $M_{*}$/$M_{\odot}$ = 0.600 $\pm$ 0.005. Within the uncertainties, these parameters agree with those derived independently from the follow-up spectral fitting, Gaia XP spectrum,  Gaia color-magnitude diagram, and SED fitting. The asteroseismological distance inferred from the optimal model differs only 1.45\% from the trigonometric parallax distance provided by Gaia. The optimal model identifies 288.417\,s and 400.640\,s as modes l = 1 and period 308.435\,s as an l = 2 mode.

Figure\,\ref{fig:results_comp} illustrates the position of WFST J0530 within the pulsation instability strip, as determined by various methods. The results from asteroseismology, specteal fitting, Gaia XP spectrum, and Gaia color-magnitude diagram are represented by different symbols. The theoretical blue and red edges depicted in the figure originate from \cite{2015ApJ...809..148T}. The gray data points in the figure, restricted by 10,600 $\leq$ $T_{\rm eff}$ $\leq$ 12,600 and 7.5 $\leq$ log\,$g$ $\leq$ 9.1, are chosen from the spectral data of DAV stars reported by \cite{2019A&ARv..27....7C}, \cite{2020AJ....160..252V}, \cite{2021ApJ...912..125G}, \cite{2022MNRAS.511.1574R}, and \cite{2025ApJ...984..112R}. The pulsation instability strip is considered to be a pure region \citep{2004ApJ...600..404B,2009BAAA...52..317C,2017EPJWC.15201011K}. However, some DAV stars are observed to lie outside this strip, which may caused by observational errors or uncertainties in measuring $T_{\rm eff}$ and log\,$g$ \citep{2009BAAA...52..317C}. As illustrated, all results from different independent measurements are consistent within their errors.

Some limitations are recognized in this study. The small variations of the dim DAV WFST J0530, $\sim$ 0.1\,mag fluctuation at G magnitude of 19.13, poses challenges for us to obtain additional equal quality photometric data. Therefore, in order to ensure our asteroseismology result is correct, various independent methods are adopted to verify.  While these efforts have been productive, there is still scope for expanding our current research. Future research can be devoted to deepening our understanding of WFST J0530 via more photometric observations and advanced asteroseismological models.

\section*{Acknowledgements}

The Wide Field Survey Telescope (WFST) is a joint facility of the University of Science and Technology of China, Purple Mountain Observatory.

J.C.G acknowledges the National Natural Science Foundation of China (NSFC) under grants 12203006, Young Scholar Program of the Beijing Academy of Science and Technology (25CE-YS-02).
J.L. is supported by the National Natural Science Foundation of China (NSFC; Grant Numbers 12403038), the Fundamental Research Funds for the Central Universities (Grant Numbers WK2030000089), and the Cyrus Chun Ying Tang Foundations.

This work has made use of data from the European Space Agency (ESA) mission Gaia (\url{https://www.cosmos.esa.int/gaia}), processed by the Gaia Data Processing and Analysis Consortium (DPAC;
\url{https://www.cosmos.esa.int/web/gaia/dpac/consortium}). Funding for the DPAC has been provided by national institutions, in particular the institutions participating in the Gaia Multilateral Agreement.
This research has made use of the SIMBAD database, operated at CDS, Strasbourg, France \citep{2000A&AS..143....9W}. The authors acknowledge the use of ChatGPT (OpenAI) for language editing. All scientific interpretations, results, and conclusions are the authors’ own.



\label{lastpage}

\end{document}